\documentclass[pra,preprint,showpacs]{revtex4}
\usepackage[centertags]{amsmath}
\usepackage{amsfonts}
\usepackage{amssymb}
\usepackage{amsthm} 
\usepackage{newlfont}
\usepackage{epsfig}
\usepackage{amscd}
\usepackage{graphicx}
\usepackage{epsfig}
\usepackage{color}
\usepackage{amscd}

\newcommand{\beq}{\begin{equation}}
\newcommand{\eeq}{\end{equation}}
\newcommand{\ba}{\begin{array}}
\newcommand{\ea}{\end{array}}
\newcommand{\bea}{\begin{eqnarray}}
\newcommand{\eea}{\end{eqnarray}}

\usepackage{graphicx}
\usepackage{epstopdf}
\usepackage{amsmath}

\begin{document}

\begin{center}
{\large \sc \bf { Extension of the remotely creatable  region via the
local unitary transformation on the receiver side.
}}

\vskip 15pt

{\large 
G.A. Bochkin and A.I.~Zenchuk 
}

\vskip 8pt

{\it Institute of Problems of Chemical Physics, RAS,
Chernogolovka, Moscow reg., 142432, Russia},\\
 e-mail:  bochkin.g@yandex.ru, zenchuk@itp.ac.ru 

\end{center}

%\today

\begin{abstract}
We consider the remote state creation via the homogeneous spin-1/2 chain
and show that the significant extension of the 
creatable region can be achieved using the local unitary transformation  of the so-called extended  
receiver (i.e. receiver joined with the nearest node(s)). This transformation is { most}  
effective in the models with all-node 
interactions. We consider the  model with the two-qubit
 sender, one-qubit receiver and two-qubit extended receiver.
\end{abstract}

\maketitle

%\graphicspath{{..}}
\section{Introduction}
\label{Section:Introduction}
The problem of remote state creation \cite{PBGWK2,PBGWK,DLMRKBPVZBW,XLYG,PSB} 
is an alternative to the state teleportation \cite{BBCJPW,BPMEWZ,BBMHP,YS1,YS2,ZZHE} and the state transfer problem
\cite{Bose,CDEL,ACDE,KS,GKMT,WLKGGB,CRMF,ZASO,ZASO2,ZASO3,SAOZ,KZ_2008,NJ,BK,C,QWL,B,JKSS,SO,BB,SJBB,LS,BBVB,Z_2014,BZ_2015}.
All of them  are aimed at the proper way of the information transfer \cite{YBB,Z_2012,PS} from the sender to the receiver. 
In most experiments  the information carriers are photons \cite{ZZHE,BPMEWZ,BBMHP,PBGWK,DLMRKBPVZBW,XLYG}.
However, the  spin chain   as a
transmission line between the sender and receiver is popular in numerical simulations, see 
\cite{Bose,CDEL,ACDE,KS,GKMT,WLKGGB,
CRMF,ZASO,ZASO2,ZASO3,SAOZ,KZ_2008,BK,C,QWL}. 

In the recent paper \cite{BZ_2015} we give the detailed description of the remote state creation in long homogeneous 
chains as  the map
(control parameters) $\to$ (creatable parameters). Here, we call 
the arbitrary parameters of the sender's initial state the control parameters, 
while the creatable parameters  are the parameters of the receiver's state 
(which are eigenvalue-eigenvector parameters in that paper).
As a characteristic of the state creation effectivity, the interval of the largest  creatable eigenvalue  
was proposed. The   critical length $N_c=34$ was found such that 
any allowed eigenvalues can be created, i.e., the largest eigenvalue  can take any value from the interval
$\frac{1}{2}\le \lambda_{max} \le 1$. It was shown that the creatable region 
of the receiver's state space { (i.e., the subregion of the receiver's state space which 
can be remotely created by varying the control parameters)} shrinks to
$\lambda_{max} =1$ with an increase in 
the length of the homogeneous spin chain. 

\begin{figure*}
   \epsfig{file=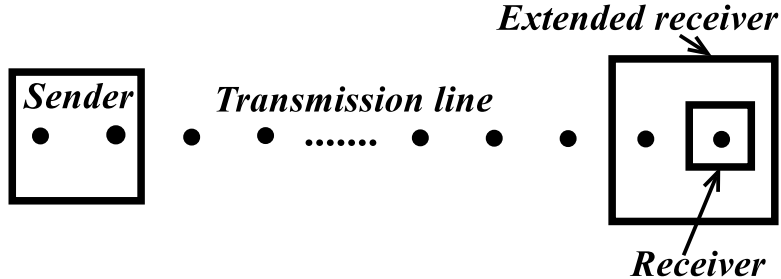,
  scale=0.5
   ,angle=0%270
}

\caption{The communication line with the extended receiver. 
Using  the {  optimized} local unitary transformation of the extended receiver 
we increase the critical length $N_c$ and thus extend the creatable region in comparison with the 
model without { the above} local transformation of the extended receiver. 
} 
  \label{Fig:comm} 
\end{figure*}

An additional  simple way to extend  the  creatable region  
{  (and thus to (partially) compensate the above mentioned shrinking of this region in long communication lines)} 
could be a local unitary 
transformation of the receiver. However this transformation can not change   
the eigenvalues { (which are part of the creatable parameters)} of the receiver state. Nevertheless,
the receiver's eigenvalue 
can be changed  by a local transformation  applied  to the so-called extended receiver 
involving the receiver  as a 
subsystem, Fig. \ref{Fig:comm}. 
Further  numerical simulations {  with the one-node receiver}  (justified by the theoretical arguments) show that 
this procedure {  is  most effective in chains governed by the Hamiltonian with all-node interactions rather then with
nearest-neighbor ones.}
As a result, we manage to significantly extend  the creatable region and increase the 
mentioned above critical length  up to $N_c=109$.

All in all,  we consider the communication line based on the homogeneous spin chain with all-node interactions 
consisting of the following parts, Fig. \ref{Fig:comm}.
\begin{enumerate}
\item
The two node sender with an arbitrary pure state whose parameters are referred to as the 
control parameters (the first and the second nodes of the spin chain).
\item
The one-qubit receiver whose state-parameters are referred to as the creatable parameters (the last node of the chain). 
\item
The two-node extended receiver consisting of the two last nodes of the chain (involving the receiver itself). 
\item
The transmission line connecting the sender with the extended receiver.
\end{enumerate}
 
 Our purpose is to modify the remote state creation algorithm given in ref.\cite{BZ_2015} using the optimal 
 local unitary transformation of the extended receiver with the purpose of extending the creatable region.

 The paper is organized as follows.
 In Sec.\ref{Section:XY}, we specify  the interaction  Hamiltonian 
 together with the initial condition used for the remote state creation.
 Sec.\ref{Section:optV} is devoted to the optimization of the local unitary 
 transformation of the extended receiver with the purpose to obtain the largest creatable region. Numerical simulations confirming the theoretical predictions 
 are presented  in Sec.\ref{Section:Num}. General conclusions are given in Sec.\ref{Section:conclusions}.

\section{XY Hamiltonian and initial state of communication line}
\label{Section:XY}
Our model of communication line is based on  the homogeneous spin chain  
with the one-spin excitation whose  dynamics is governed by the XY-Hamiltonian
\begin{eqnarray}\label{XY}
H=\sum_{{i,j=1}\atop{j>i}}^N D_{ij} (I_{i,x} I_{j,x} + I_{i,y} I_{j,y}), \;\;D_{ij}=\frac{\gamma^2 \hbar}{r^3_{ij}},
\end{eqnarray}
where $\gamma$ is the gyromagnetic ratio, $r_{ij}$  is the distance between the $i$th 
and the $j$th spins, $I_{i,\alpha}$ 
($\alpha=x,y,z$) is the projection operator of the $i$th spin on the $\alpha$ axis, 
$D_{ij}$ is the dipole-dipole coupling constant between 
the $i$th and the 
$j$th  nodes. Below we use the dimensionless time (formally setting $ D_{12} =1$).
Obviously, this Hamiltonian commutes with the $z$-projection of the total angular momentum 
$I_z$, so that the evolution of the 
one-spin excitation can be described in the
$N+1$-dimensional basis (instead of the general $2^N$ dimensional one)
\begin{eqnarray}
|{ i}\rangle, \;\;i=0,\dots,N,
\end{eqnarray}
where $| i\rangle$, $i>0$,  denotes the state with the $i$th excited spin, 
$|{0}\rangle$ corresponds to the ground state of the spin chain with zero (by convention) eigenvalue. 

The general form of the initial state of the $N$-node chain with the one-excitation initial state of the 
two-qubit sender  reads  
\begin{equation}
|\psi_0\rangle = a_0 | 0\rangle +a_1| 1\rangle+a_2| 2\rangle \label{initstate},\;\;\sum_{i=0}^2 |a_i|^2 =1,
\end{equation}
where the real parameter $a_0$ and  the complex  parameters $a_1$, $a_2$ are given as:
\begin{eqnarray}\label{alpha}
&&
a_0=\sin\frac{\alpha_1\pi}{2},\;\;a_1=\cos\frac{\alpha_1\pi}{2}\cos\frac{\alpha_2\pi}{2} e^{2 i \pi \varphi_1}\;\;
a_2=\cos\frac{\alpha_1\pi}{2}\sin\frac{\alpha_2\pi}{2} e^{2 i \pi \varphi_2},\\\label{int}
&&
0\le \alpha_i \le 1,\;\;\;0\le \varphi_i \le 1,\;\;\;i=1,2.
\end{eqnarray}
{  Note that} formula (\ref{initstate}) means that the both extended receiver and  transmission line are in the ground state initially.

\section{Optimal local transformation of the extended receiver}
\label{Section:optV}
In this section we derive the optimal local unitary transformation of the extended receiver 
which maximizes 
the creatable region. 
For this purpose we first
find the general formula for the state of the extended receiver, Sec.\ref{Section:genst}. 
Then {  we diagonalize this state using the appropriate unitary transformation 
$V$} and show that both  non-zero 
eigenvalues  depend on the probability of the 
excitation transfer to the extended receiver, Sec.\ref{Section:ev}.
After that we maximize this excitation transfer probability optimizing 
the control parameters, Sec.\ref{Section:contrmax}. 
The unitary transformation $V$ corresponding to the optimized control parameters is
the needed unitary transformation of the extended receiver, Sec.\ref{Section:V}. {  This is the transformation 
which provides the transfer of the both nonzero eigenvalues of the extended receiver to the one-node receiver.}  
After optimization of $V$ over the time $t$ (Sec.\ref{Section:t}), 
we {  obtain}  the algorithm of 
 remote state creation in Sec.\ref{Section:alg}.
 {  It is remarkable that the optimization of the  transformation $V$ can be done using the singular
 value decomposition of some special 
 matrix $P$ of  transition amplitudes  (\ref{fPa})  which simplifies numerical simulations.}
 %Useful recommendations simplifying the numerical 
%calculation of the both optimal control parameters and appropriate unitary transformation $V$ are given 
in Sec.\ref{Section:singular}.

\subsection{General state of extended receiver}
\label{Section:genst}
As mentioned above, the state of the extended receiver is described by the density matrix 
reduced over all the nodes except the two last ones. 
Written in the basis 
\begin{eqnarray}\label{basER}
|0\rangle, \;\; 
|N-1\rangle,\;\;|N\rangle,\;\;|(N-1)N \rangle,
\end{eqnarray}
this state reads
\begin{eqnarray}\label{rhoB}
\rho_{R_{ext}}\equiv {\mbox{Tr}}_{1,2,\dots,N-2} \rho=\left(
\begin{array}{cccc}
1-|f_{N-1}|^2-|f_{N}|^2 & f_{0}f_{N-1}^* &f_{0}f_{N}^*&0\cr
f_{0}^*f_{N-1} &|f_{N-1}|^2 & f_{N-1} f_{N}^*&0\cr
f_{0}^*f_{N}&f_{N-1}^* f_{N}& |f_{N}|^2 &0\cr
0&0&0&0
\end{array}
\right).
\end{eqnarray}
In (\ref{basER}), $|(N-1)N \rangle$ means the state with the  two last excited nodes of the chain,
the trace is taken over the nodes $1,\dots,N-2$,
the star means the complex conjugate value and $f_{N-1}$, $f_N$, $f_0$ are the  {  transition}
amplitudes,
\begin{eqnarray}
f_i&=&\langle i| e^{-i H t} |\Psi_0\rangle = R_{i} e^{2 \pi i \Phi_i},\;\;i=0,\dots,N,\\\nonumber
&&
0\le \Phi_i \le 1, \;\;\; R_i \ge 0.
\end{eqnarray}
Remember the  natural constraint  
 \begin{eqnarray}
\label{constr}
|f_0|^2+
|f_N|^2 +|f_{N-1}|^2 \le 1 \;\; \Rightarrow \;\; R_0^2+ R_N^2 +R_{N-1}^2 \le 1,
\end{eqnarray}
where the equality corresponds to the pure  state transfer to the nodes of the extended receiver 
because in this case  $f_i \equiv 0 $ ($0<i<N-1$).

Since the initial state is a linear function of the control parameters $a_i$, 
the transition amplitudes 
are also  linear functions 
of these parameters:
 \begin{eqnarray}\label{NN}
f_N(t)&=&\langle N| e^{-i H t} |\Psi_0\rangle = \sum_{j=1}^2 a_j \langle N| e^{-i H_1 t} |j\rangle
=\sum_{j=1}^2 a_j p_{Nj}(t)
,\\\label{N0}
f_0(t)&=&\langle 0| e^{-i H t} |\Psi_0\rangle = a_0\equiv R_0,
\end{eqnarray}
where $p_{kj}$ are transition amplitudes:
\begin{eqnarray}\label{def_chi}
  p_{kj}(t)&=&\langle k| e^{-iH_1 t}|j\rangle = r_{kj}(t) e^{2 \pi i \chi_{kj}(t)},\;\;k,j>0,\\\nonumber
  &&
  r_{kj} \ge 0,\;\;\; 0\le \chi_{kj} \le 1.
 \end{eqnarray} 
In eq. (\ref{N0}), we use  the fact that the ground state has zero energy by convention.
 We emphasize that the  transition amplitudes represent the 
 inherent characteristics of the transmission
 line and do not depend on the control parameters of the sender's initial state.
 
\subsection{Eigenvalues of extended receiver}
\label{Section:ev}

{  The construction of the optimal local transformation of the extended receiver is based on the maximization of the 
variation intervals of the creatable eigenvalues of the density matrix $\rho_{R_{ext}}$ (\ref{rhoB}) of the 
 extended receiver.
These eigenvalues} read as follows:
\begin{eqnarray}\label{lampm}
\hat \lambda_{\pm}=\frac{1}{2}\left(1\pm \sqrt{(1-2R^2)^2 + 4 R^2 R_0^2}\right),
\end{eqnarray}
where we introduce the probability of the {  excitation} transfer to the nodes of the extended receiver 
\begin{eqnarray}\label{RNN}
 R^2\equiv |f_{N-1}|^2 + |f_N|^2= R_N^2 + R_{N-1}^2.
\end{eqnarray}
The biggest eigenvalue $\hat \lambda_+$ as a function of $R$ and $R_0$ 
varies inside of some interval 
\begin{eqnarray}
\hat \lambda_0 \le \hat \lambda_+ \le 1.
\end{eqnarray}
Thus, to obtain the largest variation interval we need to minimize  $\hat \lambda_0$ as a function of $R$ 
and $R_0$. It is simple to show that the minimum   $\hat\lambda^{min}_0$ corresponds to $R_0$ =0.
For this purpose we use the following substitution prompted by 
 constraint (\ref{constr}):
\begin{eqnarray}
R_N=\hat R_N\sqrt{1-R_0^2} ,\;\; R_{N-1}= \hat R_{N-1}\sqrt{1-R_0^2},\;\;\hat R^2=\hat R_N^2 +\hat R_{N-1}^2.
\end{eqnarray}
In terms of the new  notations, the largest eigenvalue reads
\begin{eqnarray}\label{lampm2}
\hat \lambda_{+}=\frac{1}{2}\left(1 + \sqrt{1-4(1-R_0^2)^2 \hat R^2(1-\hat R^2)}\right)
\end{eqnarray}
Calculating the derivative of $\lambda_+$ with respect to $\hat R$ we find the extremum at 
$\hat R^2 =\frac{1}{2}$:
\begin{eqnarray}
\hat \lambda_+^{min}=\frac{1}{2}\left(
1+ R_0\sqrt{2-R_0^2}
\right)
\end{eqnarray}
which is minimal at $R_0 = a_0=0$: 
\begin{eqnarray}
\hat\lambda_+^{min}|_{R_0=0} = \frac{1}{2}.
\end{eqnarray}
Note that  $R$ is a continuous function of the control parameters $a_i$, $i=1,2$, 
and $R=0$ at $a_1=a_2=0$. Consequently, if $R$ reaches some  value $R^{opt}$, then with varying $a_i$, 
we can obtain any value of $R$  inside of  the interval
\begin{eqnarray}
0\le R \le R^{opt}.
\end{eqnarray}
The largest variation interval $0\le R\le 1$ corresponds to the communication line allowing the perfect state
transfer of the excitation to the extended receiver. In this case the variation interval of 
$\lambda_+$ is also maximal,
$\frac{1}{2} \le \hat \lambda_+\le 1$.
However, in general, this variation interval is following:
\begin{eqnarray}\label{hatlamint}
&&
1-(R^{opt})^2 \le \hat \lambda_+ \le 1,\;\;
0\le (R^{opt})^2 \le \frac{1}{2},\\\nonumber
&&
\frac{1}{2} \le \hat \lambda_+\le 1,\;\;(R^{opt})^2\ge \frac{1}{2}.
\end{eqnarray}
Eq.(\ref{hatlamint}) shows that the interval of creatable  $\hat \lambda_+$ is
completely defined by the probability  of the {  excitation} transfer to the extended receiver. 
Therefore, the {  maximization} of this quantity deserves the special consideration.

\subsection{Control parameters maximizing $R$ }
\label{Section:contrmax}

The  probability  of the excitation transfer  $R$ (\ref{RNN}) is equal to 
the norm of the two-component vector 
$f=(f_{N-1}\;\; f_N)^T$ (the 
superscript $T$ means transposition),
\begin{eqnarray}\label{fPa0}
&&
f=P a,\;\;\;a= \left(
\begin{array}{c}
a_1\cr
a_2
\end{array}
\right) ,\;\;\;\\\label{fPa}
&&
P=\left(
\begin{array}{cc}
p_{(N-1)1} &p_{(N-1)2}\cr
p_{N1} & p_{N2}
\end{array}
\right)
\end{eqnarray}
Thus, the maximum $R^{opt}$ of the probability $R$ as a function of the control parameters can be found as 
\begin{eqnarray}\label{opt1}
(R^{opt})^2  = \max_{|a_1|^2+|a_2|^2=1} (| f_{N}|^2 +| f_{N-1}|^2).
\end{eqnarray}
To proceed further, 
we  write  $R^2$  in the following form 
\begin{eqnarray}\label{opt2}
R^2\equiv f^+f=a^+P^+Pa. 
\end{eqnarray}
and diagonalize the matrix $P^+P$:
\begin{eqnarray}\label{defU}
P^+P = U^+ \Lambda_0^2 U, \;\;\Lambda_0^2={\mbox{diag}}(\lambda_-^2,\lambda_+^2),\;\;
0\le \lambda_-\le\lambda_+
\end{eqnarray}
where 
\begin{eqnarray}\label{lpm}
\lambda_\pm^2 &=& \frac{1}{2} \Big(r_{(N-1)1}^2 + r_{(N-1)2}^2+ r_{N1}^2+r_{N2}^2 \pm 
\sqrt{Q}\Big)
,\\\nonumber
&&
Q=(r_{(N-1)1}^2 + r_{(N-1)2}^2+ r_{N1}^2+r_{N2}^2)^2 -\\\nonumber
&&
4 
\Big(r_{(N-1)2}^2  r_{N1}^2+r_{(N-1)1}^2  r_{N2}^2 -
\\\nonumber
&&
2 \cos (2(\chi_{(N-1)1} -\chi_{(N-1)2}-\chi_{N1}+\chi_{N2}))
r_{(N-1)1}r_{(N-1)2}r_{N1}r_{N2}
\Big),
\end{eqnarray}
So, by virtue of (\ref{defU}),  eq.(\ref{opt2}) reads
\begin{eqnarray}\label{opt22}
&&
f^+f=b^+\Lambda^2_0 b, \;\;b=U a.
\end{eqnarray}
 The mutual position of the eigenvalues in the matrix $\Lambda_0^2$ is
 taken for convenience and will be used in Sec.\ref{Section:singular}.
Now, by virtue of eq.(\ref{opt22}),  we rewrite eq.(\ref{opt1}) as follows:
\begin{eqnarray}\label{Ropt2}
(R^{opt})^2 = \max_{|b_1|^2+|b_2|^2=1} (\lambda_-^2 |b_1|^2 + \lambda_+^2 |b_2|^2).
\end{eqnarray}
Obviously, the maximal value is achieved when $b_2=1$ and $b_1=0$:
\begin{eqnarray}\label{Ropt22}
(R^{opt})^2 =\lambda_+^2 .
\end{eqnarray}
The appropriate  expression for the vector of control parameters  $a^{opt}$ follows from the relation between $a$ and $b$ 
given in the second of  eqs.(\ref{opt22}):
\begin{eqnarray}\label{a}
Ua^{opt}=\left(
\begin{array}{c}
0\cr
1
\end{array}
\right),\;\;\;
\Rightarrow
a^{opt}=U^+\left(
\begin{array}{c}
0\cr
1
\end{array}
\right)
\end{eqnarray}
Formula (\ref{a}) gives us the sender's initial state  leading
to the maximal value $R^{opt}$  of the probability $R$  at a given time instant.

%It will be shown in Sec.\ref{Section:singular}, that $(R^{opt})^2 = \lambda_+^2$.

Note that $|f_{N-1}|=0$ at the extremum point of $|f_N|$ in the case of 
nearest neighbor approximation \cite{BZ_2015}. As a result,  
$R^{max}\equiv \max\limits_{|a_1|^2+|a_2|^2=1} |f_N|$ and there is no contribution from the $(N-1)$th node of the chain. 
That is why the local transformation of the two-node extended receiver 
is not effective in the case of nearest neighbor approximation. 

\subsubsection{Explicit form of $U$}

We can also write the explicit form of $U$ (and, consequently, the explicit form of $a^{opt}$) in 
terms of the probability amplitudes $p_{ij}$.
This can be done using the  definition (\ref{defU}) written as
\begin{eqnarray}\label{eqU}
U P^+ P = \Lambda_0^2 U.
\end{eqnarray}
Let us represent  the matrix $U$ in 
terms of the parameters $\alpha^{opt}_i$ and $\varphi^{opt}_i$:
\begin{eqnarray}\label{unitU}
U=
\left(\begin{array}{cc}
\cos\frac{\alpha^{opt} \pi}{2} & - \sin\frac{\alpha^{opt} \pi}{2} e^{-2 i\varphi^{opt}} \cr
 \sin\frac{\alpha^{opt} \pi}{2} e^{2 i\varphi^{opt}}&\cos\frac{\alpha^{opt} \pi}{2} 
\end{array}
\right).
\end{eqnarray}
Substitute matrix (\ref{unitU}) into eq.(\ref{eqU}) we can solve it for 
the  parameters $\varphi^{opt}$ and $\alpha^{opt}$:
\begin{eqnarray}\label{optimal}
&&
\tan(\frac{\alpha^{opt} \pi}{2}) = \\\nonumber
&&\frac{r_{N1}^2 + r_{(N-1)1}^2- \lambda_-}{
\cos(2\pi (\phi_{12} - \chi_{(N-1)1} + \chi_{(N-1)2})) r_{(N-1)1}r_{(N-1)2}+
\cos(2 \pi(\phi_{12} - \chi_{N1} +  \chi_{N2})) r_{N1}r_{N2}
}
,
\\\nonumber
&&
\tan(2 \varphi^{opt} \pi)=\frac{
\sin(2 \pi( \chi_{(N-1)1} -  \chi_{(N-1)2})) r_{(N-1)1}r_{(N-1)2}+
\sin(2\pi( \chi_{N1} -  \chi_{N2})) r_{N1}r_{N2}}{\cos(2 \pi (\chi_{(N-1)1} - 
\chi_{(N-1)2})) r_{(N-1)1}r_{(N-1)2}+
\cos(2 \pi(\chi_{N1} -  \chi_{N2})) r_{N1}r_{N2}}.
\end{eqnarray}
Then formula (\ref{a}) with $U$ given by  (\ref{optimal}) gives us the expressions for the 
control parameters maximizing the probability $R$  of the  excitation transfer 
to the nodes of the extended receiver.

\subsection{Optimized local transformation of the extended receiver }
\label{Section:V}
In Sec.\ref{Section:contrmax} we find the values of the control parameters  for construction of $R^{opt}$. 
Namely, $a_0=0$ (or $\alpha_1=0$), $\varphi_1$ is arbitrary, while 
$\alpha_2$ and $\varphi_2$ are determined by expressions (\ref{optimal}). 
Now we write the explicit form of the local transformation diagonalizing the state of the 
extended receiver obtained for the above control parameters. 
Before the diagonalization,  density matrix (\ref{rhoB}) reads 
(we  mark the appropriate quantities with the 
superscript $opt$):
\begin{eqnarray}\label{rhoB2}
\rho^{opt}_{R_{ext}}\equiv {\mbox{Tr}}_{1,2,\dots,N-2} \rho=\left(
\begin{array}{cccc}
1-(R^{opt})^2  & 0&0&0\cr
0 &|f^{opt}_{N-1}|^2 & f^{opt}_{N-1} (f^{opt}_{N})^*&0\cr
0&(f^{opt}_{N-1})^* f^{opt}_{N}& |f^{opt}_{N}|^2 &0\cr
0&0&0&0
\end{array}
\right).
\end{eqnarray}
It is remarkable that  the central nonzero $2\times 2$ block of the density matrix $\rho^{opt}_{R_{ext}}$ 
can be factorized as
\begin{eqnarray}
\left(\begin{array}{cc}
|f^{opt}_{N-1}|^2 & f^{opt}_{N-1} (f^{opt}_{N})^*\cr
(f^{opt}_{N-1})^* f^{opt}_{N}& |f^{opt}_{N}|^2
\end{array}\right) =f f^+.
\end{eqnarray}
It is clear that this block can be diagonalized by the matrix $V_0$ of the following form:
\begin{eqnarray}\label{tran1}
V_0=\frac{1}{R^{opt}}\left(
\begin{array}{cc}
\displaystyle
f^{opt}_N & -f^{opt}_{N-1} \cr
(f^{opt}_{N-1})^* & (f^{opt}_N)^*
\end{array}
\right)
\end{eqnarray}
with the eigenvalue matrix 
\begin{eqnarray}\label{Lambda}
\Lambda_b={\mbox{diag}}(0,(R^{opt})^2),
\end{eqnarray}
so that we can write
\begin{eqnarray}\label{extd}
 \rho^{opt}_{R_{ext}} = V^+\Lambda V,
\end{eqnarray}
where
\begin{eqnarray}\label{Vopt}
&&
V={\mbox{diag}}(1,V_0,1),\;\; \\\label{Lamopt}
&&
\Lambda={\mbox{diag}}(1-(R^{opt})^2,\Lambda_b,0).
\end{eqnarray}
Consequently, applying the unitary transformation 
$V$ (\ref{Vopt}) to $\rho^{opt}_{R_{ext}}$ we obtain the diagonal density matrix 
\begin{eqnarray}\label{diagRho}
\tilde\rho^{opt}_{R_{ext}} =
V\rho^{opt}_{R_{ext}} V^+=  \Lambda.
\end{eqnarray}
The transformation (\ref{Vopt}) with $V_0$ from (\ref{tran1}) 
is the needed local unitary  transformation of the extended receiver.

\subsubsection{The optimized state of one-qubit receiver}

To obtain the optimized state of the receiver, we reduce the density matrix (\ref{diagRho}) 
 to the state of the last node using the basis (\ref{basER}). {  
 Owing to the mutual positions of the 
eigenvalues in the diagonal matrix (\ref{Lamopt}), this state reads:}
\begin{eqnarray}\label{rhoNopt}
\rho^{opt}_R={\mbox{diag}}(1-(R^{opt})^2,(R^{opt})^2).
\end{eqnarray}
 Thus the both non-zero eigenvalues are 
transferred from the extended receiver to the receiver itself.

\subsection{Singular value  decomposition of $P$ in terms of  matrices  $V^+_0$, $U$ and $\Lambda_0$}
\label{Section:singular}

It is remarkable that the matrices   $V_0$, $U$ and $\Lambda_0$  can be given another meaning. In fact, 
the central $2\times 2$ block of 
eq.(\ref{diagRho}) by virtue of eqs.(\ref{tran1},\ref{Lambda},\ref{Vopt},\ref{Lamopt}) yields
\begin{eqnarray}\label{VffpV0}
V_0 f f^+V_0^+ = (R^{opt})^2 \left(
\begin{array}{cc}
0&0\cr
0&1
\end{array}
\right).
\end{eqnarray}
On the other hand, eq.(\ref{opt2}) by virtue of eq.(\ref{defU}) can be written in the form
\begin{eqnarray}\label{fpf}
f^+ f = a^+ U^+ \Lambda_0 \tilde V \tilde V^+ \Lambda_0 U a,
\end{eqnarray}
where $\tilde V$ is some unitary matrix.
Now we can formally split  eq. (\ref{fpf}) into equation for  $f$ 
\begin{eqnarray}\label{fpff0}
f &=&  \tilde V^+ \Lambda_0 U a \;\;\Rightarrow \;\;\\\label{fpff}
 \label{tVf}
\tilde V f &=& \Lambda_0 U  a \stackrel{(\ref{a})}{=} \left(\begin{array}{c}
 0\cr
 \lambda_+ 
 \end{array}\right)
\end{eqnarray}
and its Hermitian conjugate.
Multiplying eq.(\ref{tVf}) by its Hermitian conjugation from the right we obtain
\begin{eqnarray}\label{VffpV}
\tilde Vff^+ \tilde V^+ = \lambda_+^2 \left(
\begin{array}{cc}
0&0\cr
0&1
\end{array}
\right).
\end{eqnarray}
Comparison of  eqs. (\ref{VffpV0}) and (\ref{VffpV}) {  prompts us} to identify   
\begin{eqnarray}\label{rel}
\tilde V= V_0 ,\;\; R_{opt}^2 =\lambda_+^2.
\end{eqnarray}
Comparing eq.(\ref{fpff}) with eq.(\ref{fPa0}) for $f$ by virtue of eqs.(\ref{rel})  we conclude  that  
\begin{eqnarray}
P=V^+_0\Lambda_0 U,
\end{eqnarray}
i.e., the  matrices $V^+_0$, $U$ and $\Lambda$ constructed in  Secs. \ref{Section:contrmax} and 
\ref{Section:V} represent the 
singular value decomposition  of the matrix $P$. 
This fact allows us to simplify the  algorithm  of the numerical  construction and time-optimization 
of  the probability $R^{opt}$ together with the unitary transformations 
 $V$ and $U$. This algorithm reads as follows.
\begin{enumerate}
\item
 Calculate the matrix $P$
as a function of the  time $t$ for the given Hamiltonian governing the spin dynamics and calculate its largest singular
value $\lambda_+$ as a 
function of the time $t$. 
\item \label{st3}
Find the time instant $t_0$ maximizing the  largest singular value  of the matrix $P$.
This maximal singular value gives the maximized  probability 
\begin{eqnarray}\label{Rmax}
(R^{(max)})^2\equiv (R^{(opt)})^2|_{t=t_0} = (\lambda_+)^2|_{t=t_0}
.
\end{eqnarray}
\item
Construct the  singular decomposition of $P$ at the time instant $t=t_0$ obtaining the matrices  $U^{max}$ 
(the optimized unitary transformation of the sender) and 
 $V^{max}$ (the optimized unitary transformation of the extended receiver):
\begin{eqnarray}\label{max}
V^{max} = V|_{t=t_0},\;\;\; U^{max} = U|_{t=t_0}.
\end{eqnarray}
\end{enumerate}

%Notice that eq.(\ref{lpm}) yields the  needed quantity $R_{opt}=\lambda_+$. 
%It is interesting to note that the  matrix $U$ used above does not appear in the suxceeding calculations 
%of the creatable region.

Especially important in the above algorithm is the time-optimization of the probability $R^{opt}$ in no.\ref{st3}
which is  given the  special consideration in the next paragraph.

\subsubsection{Time-maximization of  probability $(R^{opt})^2$ }
\label{Section:t}
\begin{figure*}
  \epsfig{file=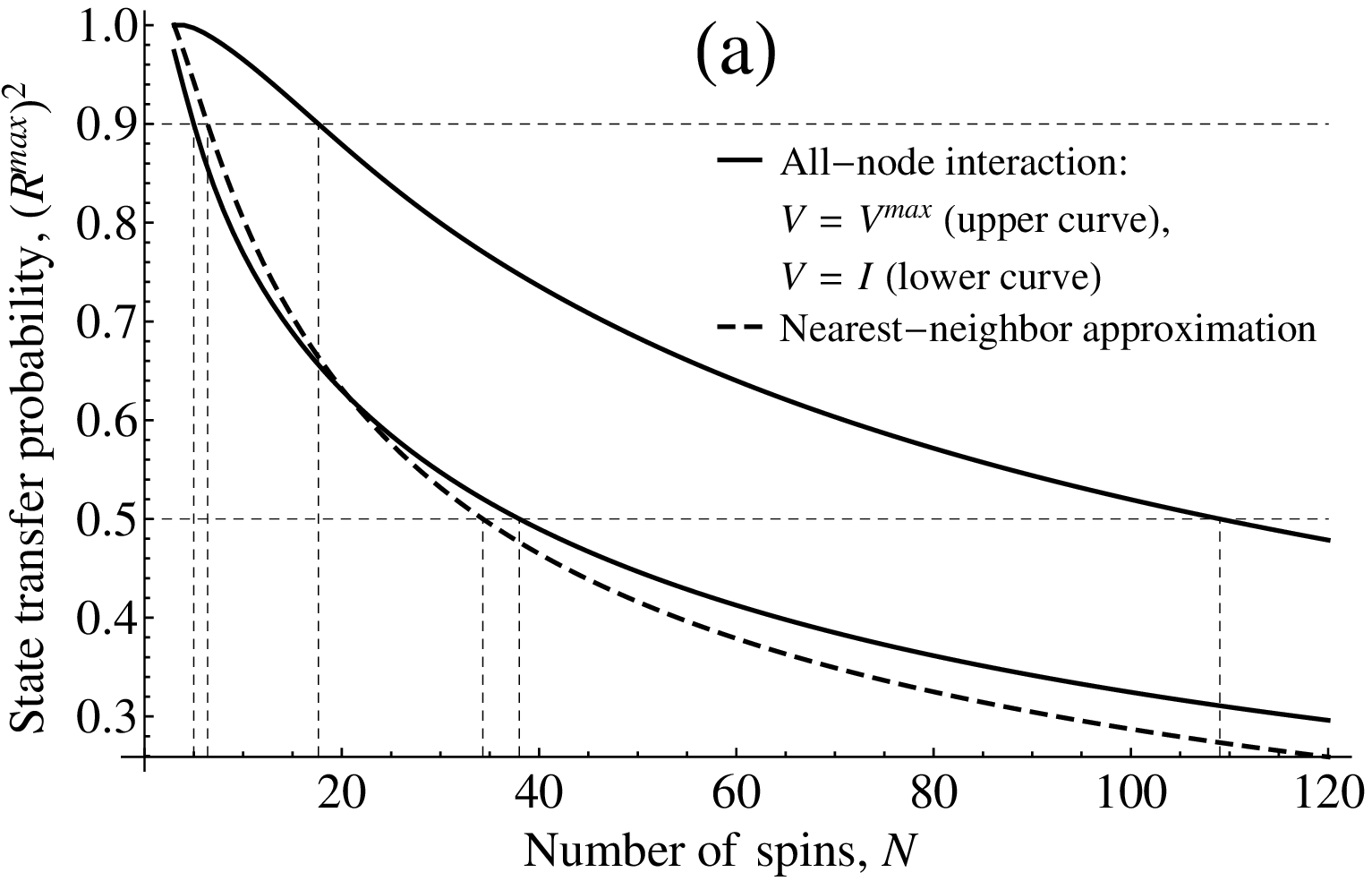,
  scale=0.5
   ,angle=0%270
} \epsfig{file=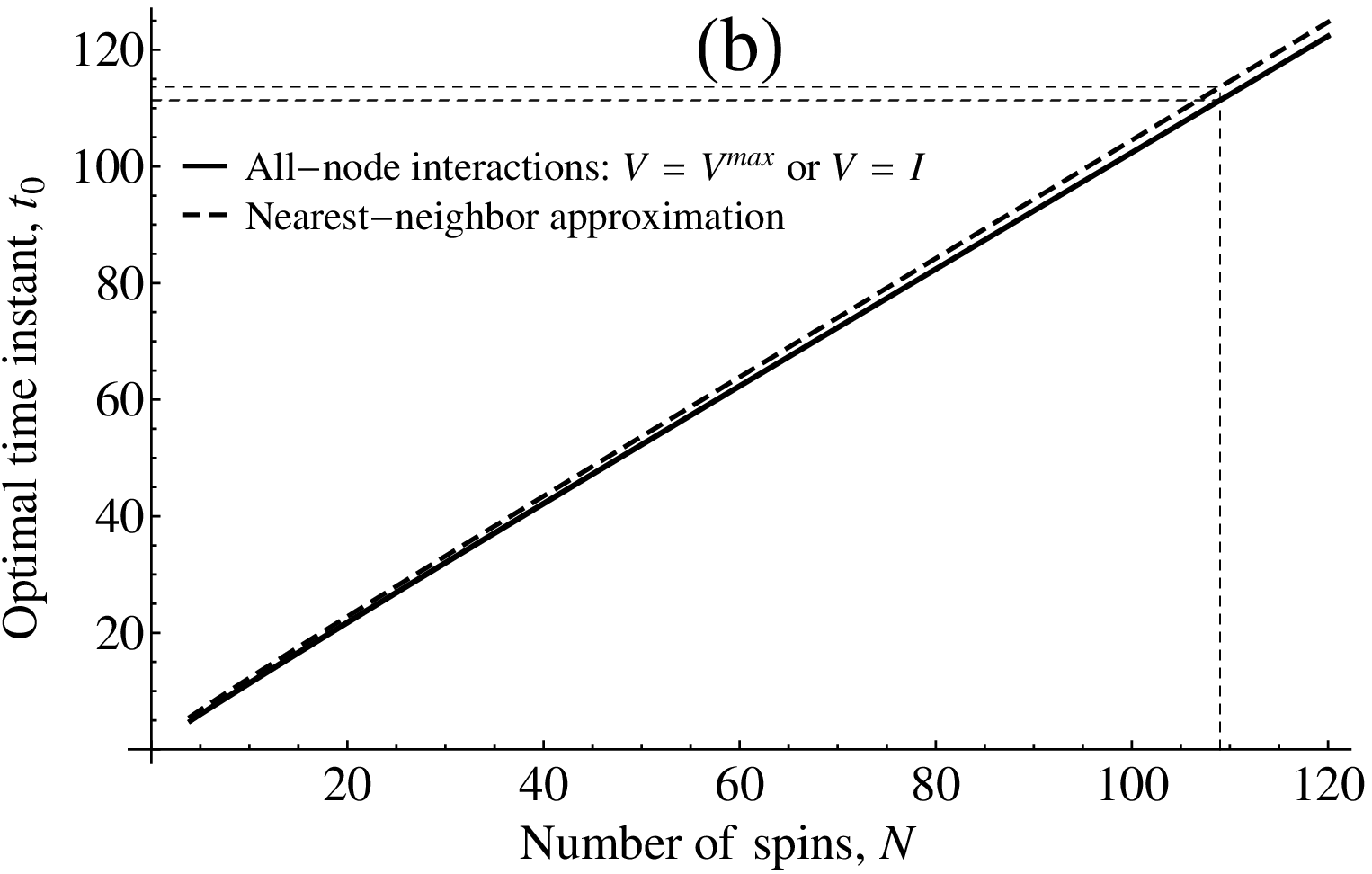,
  scale=0.5
   ,angle=0%270
}
\caption{ (a)
The maximal excitation  transfer probability $(R^{max})^2$  
and (b) the corresponding time instant $t_0$  as 
functions of the chain length $N$ 
in different models based on the  homogeneous spin-1/2 chain with XY-Hamiltonian:
 nearest neighbor approximation (the dash-line),
 all-node interactions 
without the local transformation of the extended receiver (the lower solid line),
 all-node interactions involving the  optimal   local transformation of the extended receiver 
 (the upper solid line). 
 The upper and lower horizontal
 dot-lines indicate, respectively,  the 
 lower limit of the high-probability state transfer ($(R^{max})^2=0.9$) 
 and the minimal  value of $(R^{max})^2=\frac{1}{2}$
 providing the creation of any eigenvalue ($\frac{1}{2} \le \lambda_{max} \le 1$) in the receiver's state. 
 In all cases $t_0$ is essentially  a linear function of $N$, note that  the line corresponding 
 to the model with all-node interactions involving the  local transformation $V$ is a bit {  below}
 the line corresponding to the model 
 without $V$ (although these two lines are indistinguishable in the figure). 
} 
  \label{Fig:P} 
\end{figure*}

In this subsection we give some remarks regarding the maximization of the largest singular value $\lambda_+$  
(or the probability $(R^{opt})^2$)
as a function of time. 
The probability  $(R^{opt})^2$ is an oscillating function of the time $t$ with the 
well defined first maximum \cite{BZ_2015}. 
The value of this maximum $ (R^{max})^2$ together with the corresponding time instant $t_{0}$ 
as functions of the chain length are shown in Fig.\ref{Fig:P}.
For comparison,   $(R^{max})^2$ as function of $N$ 
for the state creation without the local transformation of the extended receiver 
 is shown  for both nearest-neighbor approximation (the dash-line, $(R^{max})^2 = |f_N|^2$ in this case) 
and all-node interaction
(the lower solid line). 
We see that the high probability state transfer ($(R^{max})^2\ge 0.9$)
is possible if $N\le 6$,   $N\le 4$  and $N\le 17$ in the models, respectively, 
with   nearest-neighbor interactions, with all-node interactions without the
optimized unitary transformation $V$ and  involving  this transformation.

Another parameter indicated in Fig.\ref{Fig:P} is the critical length $N_c$ such that any 
eigenvalue can be created in the receiver if $N\le N_c$. We see that  
the excitation transfer probability exceeds the critical value $R^2_c=\frac{1}{2}$  
if $N\le 34$,  $N\le 37$ and $N\le 109$ in the models, respectively,
with   nearest-neighbor interactions, with all-node interactions without the 
optimized unitary transformation $V$ and  involving  this transformation.

\subsection{The algorithm of remote state creation. Analysis of  creatable region}
\label{Section:alg}

As the main result of this section we formulate the complete algorithm of the remote state creation.  
\begin{enumerate}
\item
Construct the optimized unitary transformations $U^{max}$ and  $V^{max}$ and the optimal time instant $t_0$ 
using the algorithm in Sec.\ref{Section:singular}.
\item \label{st2}
Create  the initial   state (\ref{initstate}) of the whole chain 
(i.e. the one-excitation pure state of the sender and the 
ground state of the rest of the chain).
\item
Apply the unitary transformation $U^{max}$ (\ref{max}) to the sender.  
\item
Switch on  the evolution  of the spin chain. 
\item
 Apply  the local unitary transformation $V^{max}$
to  the  extended receiver at the time instant $t_{0}$. 
\item
Determine the state of the receiver at the time instant $t=t_0$ 
as the trace of the whole density matrix over the all 
nodes except the receiver's node. The resulting density matrix reads as follows 
(we use the basis $|0\rangle$, $|N\rangle$):
\begin{eqnarray}\label{z}
 &&
 \rho_R={\mbox{Tr}}_{N-1} \left[V^{max} \rho_{R_{ext}} (V^{max})^+\right]= \left(\begin{array}{cc}
1-|z|^2 & f_0 z \cr
f_0z^* & |z|^2
\end{array}
\right), \;\;\;\\\label{zexpl}
&&z= \frac{1}{R^{max}} (f_{N}^* f^{max}_{N} +f_{N-1}^* f^{max}_{N-1})=R_z e^{2 i \Phi_z \pi}.
\end{eqnarray}
\end{enumerate}
This matrix coincides with $\rho^{opt}_R$ (\ref{rhoNopt}) if we use optimized initial state (\ref{a}) with $f_0=0$
at the step no.\ref{st2}.
The function $z$ in eq.(\ref{z}) is nothing but the transition amplitude to the last node 
(compare with ref. \cite{BZ_2015}) after the  evolution followed by the  local optimal transformation 
$V^{max}$ (don't mix $z$ with $f_N$!). We see that the probability of the {  excitation} transfer to the last node
$|z|^2$ reaches its maximal value 
$|z_{max}|^2 = (R^{max})^2$  for the optimal  initial state (\ref{a}) { 
and is the sum of probabilities $|f_{N-1}^{max}|^2$ and 
$|f_N^{max}|^2$. The latter statement is   the consequence of the optimizing transformation $V^{max}$. 
Without this transformation, 
we would have just  $|z_{max}|^2 = |f_N^{max}|^2$}. Thus, again, 
the probability of the excitation transfer to the receiver  of the communication line is the 
parameter responsible for the area of the creatable region in the state-space of the 
receiver \cite{BZ_2015}. We emphasize that our model allows
us to increase the length of the high probability ($\ge 90\%$) state transfer through the homogeneous spin chain 
from $N=6$ nodes (nearest neighbor approximation) to $N=17$, see Fig.\ref{Fig:P}.  

Analyzing  the creatable region we 
follow ref.\cite{BZ_2015} and  use the  eigenvalue-eigenvector parametrization of the receiver state:
\begin{eqnarray}\label{rhoULU}
\rho^B= U^B \Lambda^B (U^B)^+,
\end{eqnarray}
where 
$\Lambda^B$ is the diagonal matrix of eigenvalues and $U^B$ is the matrix of eigenvectors: 
\begin{eqnarray}
\label{Lambda2}
&&
\Lambda^B={\mbox{diag}}(\lambda,1-\lambda),\\\label{U}
&&
U^B=
\left(\begin{array}{cc}
\cos \frac{\beta_1 \pi}{2} & -e^{-2 i \beta_2\pi} \sin \frac{\beta_1\pi}{2} \cr
e^{2 i \beta_2\pi} \sin \frac{\beta_1\pi}{2}  & \cos \frac{\beta_1\pi}{2}
\end{array}\right).
\end{eqnarray}
In the ideal case, varying  $\lambda$ and $\beta_i$ ($i=1,2$)  inside of the intervals 
\begin{eqnarray}
\label{lamint}
&&
\frac{1}{2} \le \lambda \le 1, \\\label{betint}
&&
0\le \beta_i\le 1,\;\;i=1,2,
\end{eqnarray}
we can create the whole state-space of the receiver.
However, the parameters $\lambda$ and $\beta_i$ are not  arbitrary because  they depend on the control parameters
via the functions $R_0$, $R_z$ and $\Phi_z$ in accordance with the formulas \cite{BZ_2015}:
\begin{eqnarray}\label{lam}
\lambda&=&\frac{1}{2}
\left(
1+\sqrt{(1-2 R_z^2)^2 + 4 R_z^2 R_0^2 }
\right),\\
\cos\beta_1 \pi &=&\frac{1-2R_z^2}{\sqrt{(1-2R_z^2)^2 +  4 R_z^2 R_0^2}},\;\;\Rightarrow
\\\label{arccosbet1}\label{cosbet1}
\beta_1 \pi &=&\arccos\frac{1-2R_z^2}{\sqrt{(1-2R_z^2)^2 +  4 R_z^2 R_0^2}},
\\\label{beta2}
\beta_2 &=&\Phi_z.
\end{eqnarray}
As a result, the variation intervals of the creatable parameters $\lambda$ and $\beta_1$ become restricted so that the creatable 
region does not cover the whole state space of the receiver. 
On the contrary, any value of  $\beta_2$ can be constructed by the proper choice of the phases 
$\varphi_i$, $i=1,2$ in the initial state (\ref{initstate}) \cite{BZ_2015}.
This conclusion follows from the explicit expression for $z$ 
in (\ref{zexpl}).
Therefore,  below we consider the simplified map 
\begin{eqnarray}\label{map}
(\alpha_1,\alpha_2) \to (\lambda,\beta_1).
\end{eqnarray}

\section{Numerical simulations}
\label{Section:Num}
\begin{figure*}
\epsfig{file=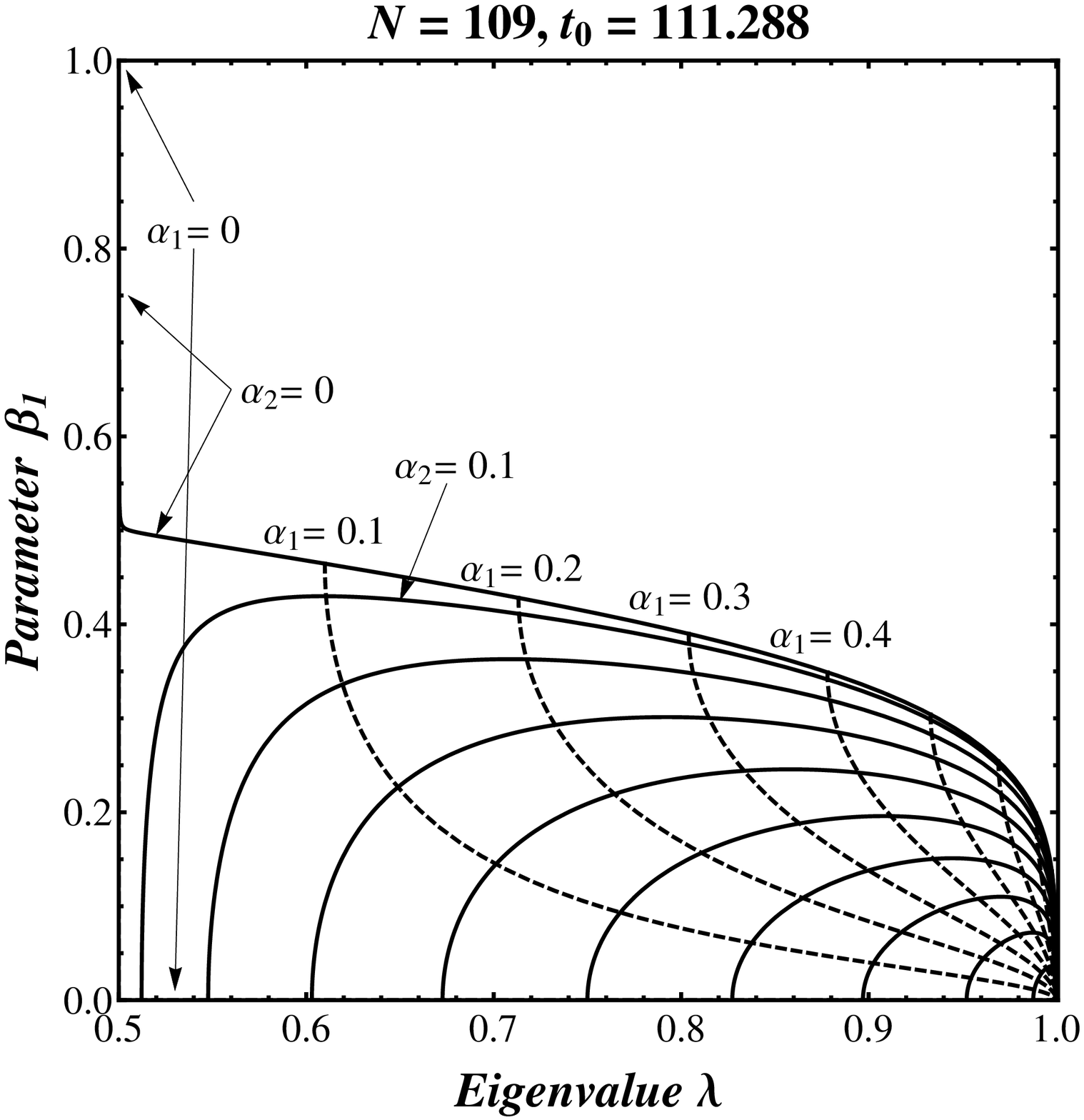,
  scale=0.36
   ,angle=0%270
}
\epsfig{file=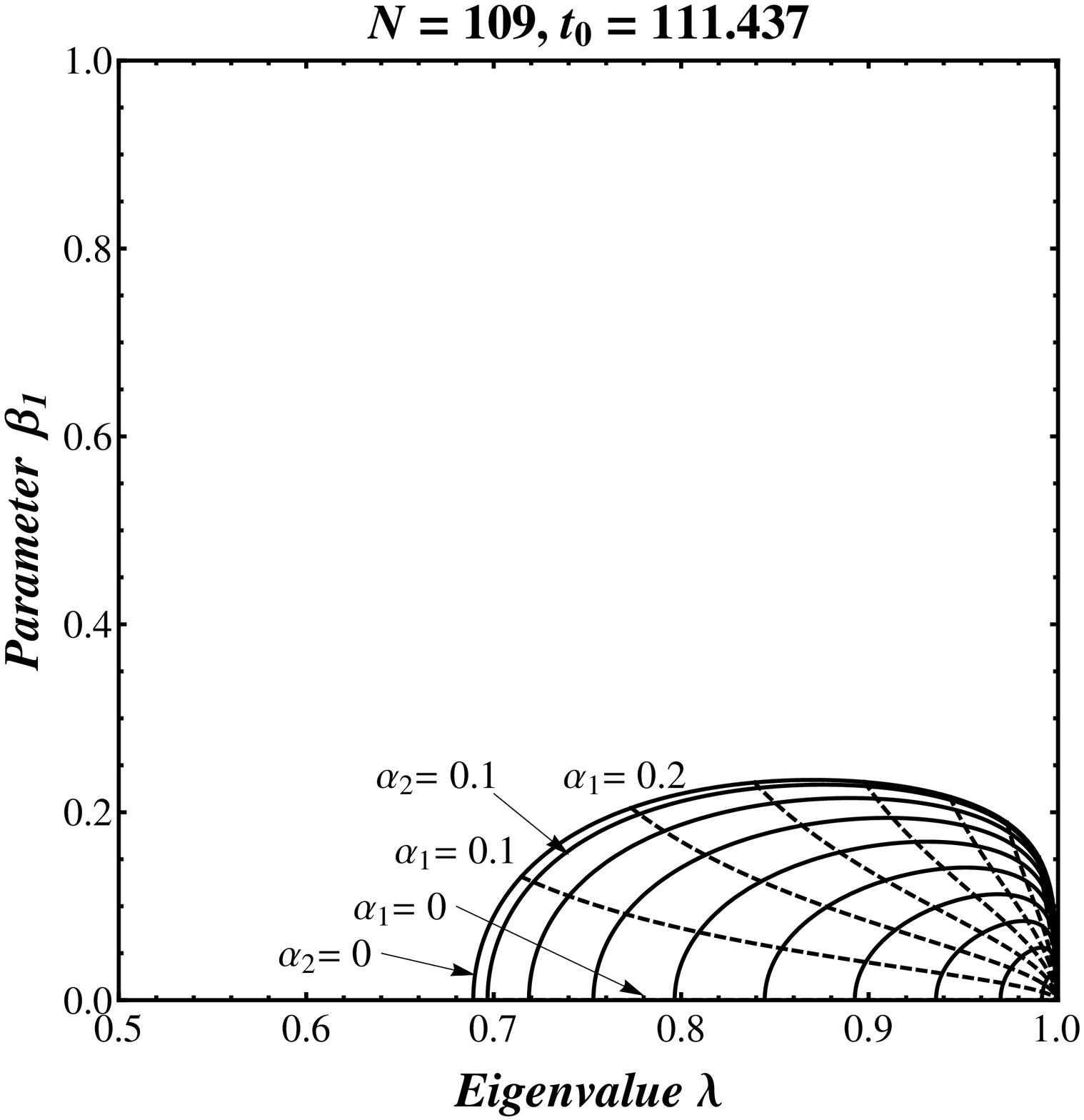,
  scale=0.36
   ,angle=0%270
}
\epsfig{file=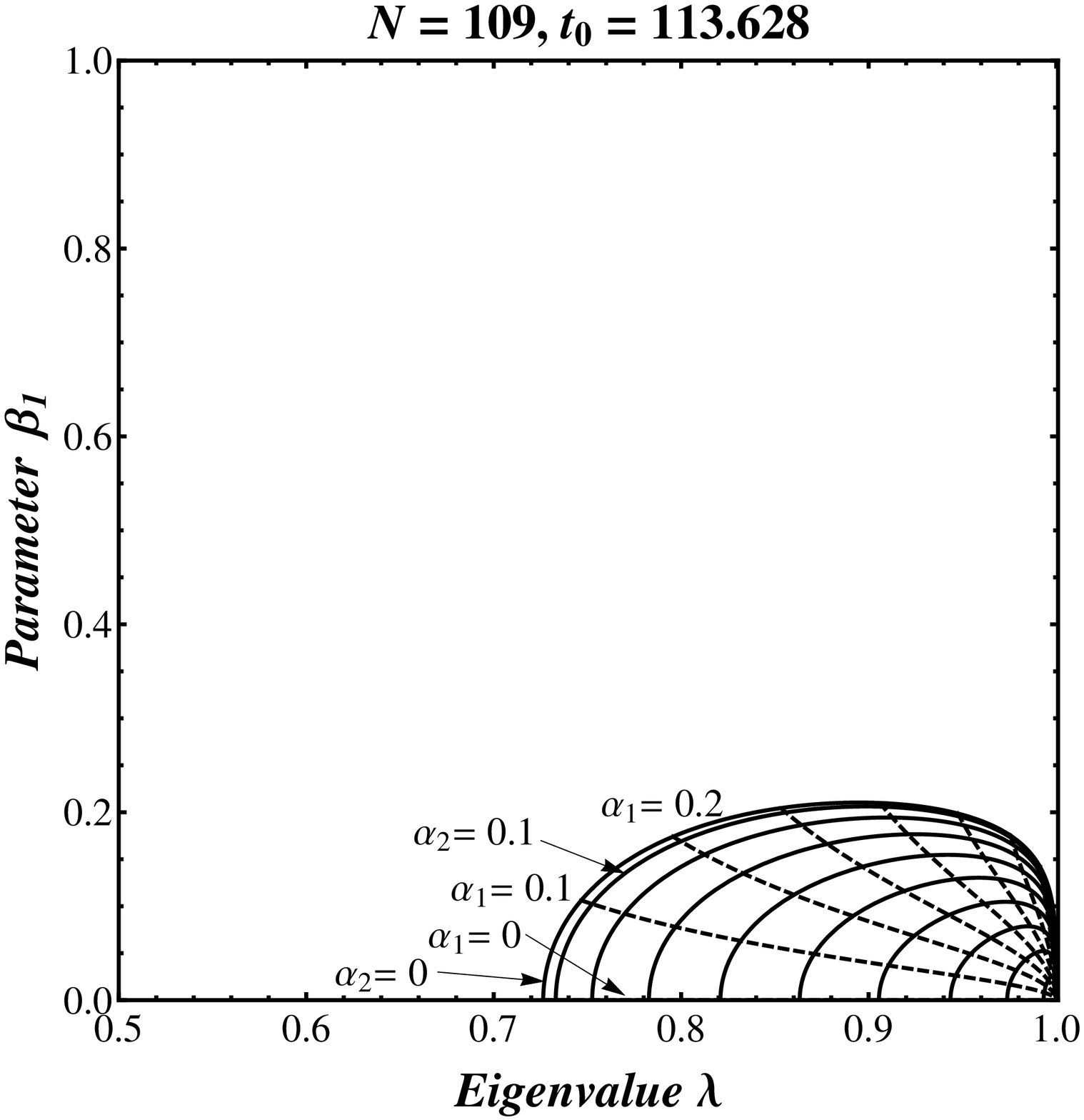,
  scale=0.36
   ,angle=0%270
}
\caption{The creatable regions in the state-space of the receiver for the different models based on the
homogeneous spin-1/2 chain and XY-Hamiltonian. (a)
The model with all-node interactions involving the  optimal   local transformation $V^{max}$ of the extended receiver;
 (b)The model with
 all-node interactions 
without the local transformation $V^{max}$  of the extended receiver;
(c) The model with nearest neighbor interactions.
Solid- and dash-lines correspond to { $\alpha_1=const$ and $\alpha_2=const$} respectively; the interval between 
the neighboring lines is $0.1$
}
\label{Fig:Num}
\end{figure*}

Now we apply the algorithm proposed in Sec.\ref{Section:alg} {  to the} numerical  study of  map (\ref{map})
 in the case of   spin chain of the critical length $N_c=109$. In   Fig.\ref{Fig:Num}, we collect the results of such simulations 
 for the different models shown  
in Fig.\ref{Fig:P}: 
the model with  all-node interactions involving the   optimized   local transformation 
 of the extended receiver ($V=V^{max}$), Fig.\ref{Fig:Num}a;
the model with all-node interactions without the  optimized   local transformation of the extended receiver
($V$ equals the identity 
matrix $I$), 
Fig.\ref{Fig:Num}b;
 the model with  nearest neighbor interactions, Fig.\ref{Fig:Num}c.
 We see that  using the all-node interaction without optimized local transformation we can only slightly extend 
 the creatable region (compare Figs. \ref{Fig:Num}b and \ref{Fig:Num}c), while the optimized transformation $V^{max}$ 
 allows us to significantly extend it, see Fig.\ref{Fig:Num}a. 
 Results of our numerical simulations confirm the theoretical predictions of Sec.\ref{Section:optV}
 regarding the extension of the creatable region.

 \section{Conclusions}
 \label{Section:conclusions}
 In this paper we show that an effective method of increasing the both   distance of the high probability state transfer and 
 creatable 
 region is the specially constructed local unitary transformation applied at the receiver side of the chain.
 This local transformation must involve not only the nodes of the receiver itself, but also some nodes from the 
 close neighborhood.
 In our case of the one-node receiver we  involve {  only the}  one additional node which 
 (together with the node of the receiver) form the two-node  extended receiver. 
 We emphasize that   this procedure {  (which uses the two-qubit extended  receiver)} is not effective  in the case of nearest neighbor approximation and  becomes useful
 if the   spin dynamics is  governed by the all node interaction Hamiltonian, which, obviously, 
 is more natural   in the case of dipole-dipole interactions. 

  As a result, we increase 
  the distance of the high probability ($\ge 90 \% $) state transfer 
  from $N=6$ (nearest neighbor approximation) to  $N=17$.
  The chain length allowing us to create any eigenvalue of the receiver is increased 
  from $N_c=34$
  (nearest neighbor approximation) to 
  $N_c=109$, as shown in Fig.\ref{Fig:P}. As a consequence, the creatable region is also extended.

 The algorithm  constructing the optimized  unitary transformation of the extended receiver $V^{max}$ is 
 described  in Sec.\ref{Section:optV} in detail. Doing this we also  obtain the initial 
 sender's state ($a^{opt}$ in eq.(\ref{a}) with $U=U^{max}$)
 maximizing the excitation transfer probability. 
 It is remarkable that the optimized 
 unitary transformation of the extended receiver  $V^{max}$ can be constructed in terms  of the 
 transition amplitudes  $p_{ij}$ ($i=1,2$, $j=N-1,N$) between the nodes of the sender and extended 
 receiver (amplitudes $p_{ij}$  are  inherent 
 characteristics of the communication line). More exactly, this transformation can be obtained from the singular 
 value decomposition of the 
 matrix $P$ (\ref{fPa}) whose elements are  the above probabilities. This observation is very useful for the numerical 
 simulation of the  remote state 
 creation. Results of such simulations are shown in Fig.\ref{Fig:Num}. 
 
 The work is partially supported by RFBR (grant 15-07-07928). A.I.Z. is 
 partially supported by DAAD (the Funding program 
 ''Research Stays for University Academics and Scientists'', 2015
(50015559))

\end{document}